\documentclass[preprint,amsmath,amssymb,amsfonts,aps,prd]{revtex4-1}

\usepackage[dvips]{graphicx}

\newcommand*{\I}{\mathrm{i}} 
\newcommand*{\E}{\mathrm{e}} 
\newcommand*{\GAMMA}{\gamma_{E}} 
\newcommand*{\alphas}{\alpha_{S}} 
\newcommand*{\mur}{\mu_{R}} 
\newcommand*{\muf}{\mu_{F}} 

\newcommand*{\Refcite}[1]{Ref.~[\onlinecite{#1}]}
\newcommand*{\Eqtextref}[1]{Eq.~(\ref{#1})}
\newcommand*{\Figref}[1]{Fig.~\ref{#1}}
\newcommand*{\Tabref}[1]{Table~\ref{#1}}
\newcommand*{\Secref}[1]{Sec.~\ref{#1}}

\begin{document}
\title{An Enhanced Threshold Resummation Formalism for Lepton Pair Production and Its Effects in the Determination of Parton Distribution Functions}
\author{David Westmark}
\affiliation{Tallahassee Community College, 444 Appleyard Drive, Tallahassee, FL 32304}
\author{J. F. Owens}
\affiliation{Florida State University, Tallahassee, Florida 32306-4350}

\begin{abstract}
We demonstrate that theoretical predictions using current resummation techniques for the lepton pair production (LPP) rapidity and $x_F$ distributions can be inconsistent with data in high rapidity and $x_F$ kinematic regions by observing their effect in global fits of parton distribution functions (PDFs). We present an enhanced resummation technique for the LPP rapidity and $x_F$ distributions that agrees with LPP data. The enhanced resummation method is used in conjunction with threshold resummation in deep inelastic scattering (DIS) to perform two global fits of PDFs using the minimal and Borel prescriptions. The results are analyzed to determine the effect of threshold resummation on global fits of PDFs.
\end{abstract}

\maketitle

\section{\label{secIntro}Introduction}
It is known that the logarithms associated with the threshold kinematic region can be large enough to potentially spoil a perturbative series. A technique to resum these large logarithms to all orders in perturbation theory, known as threshold resummation, was developed some time ago~\cite{Sterman1987a,Catani1989}. Threshold resummation can be used to improve the accuracy of calculations that require a high degree of precision and has become an important tool in QCD~\cite{Catani2007}.

The effects of threshold resummation have been studied extensively in processes such as deep inelastic scattering (DIS) and lepton pair production (LPP)~\cite{Sterman1987a,Catani1989,Vogt2001a,Bolzoni2006,Mukherjee2006}. Theoretical predictions of such processes with hadrons in the initial state require a knowledge of the parton content of the hadrons, which is described using parton distribution functions (PDFs). Recent global fits of PDFs include DIS data at high parton energy fraction $x$ and low squared transfer momentum $Q^2$ in order to provide strong constraints on the large $x$ PDFs~\cite{Alekhin2010,Alekhin2012,Accardi2010}, which is where threshold resummation is known to be relevant~\cite{Sterman2000}.

The large $x$ PDFs are important to the calculation of observables describing some processes in proton-proton collisions, as might be seen at the LHC. An inclusive cross section for a process at center of mass energy $\sqrt{s}$ that produces a final state of mass $M$ and rapidity $Y$ depends on PDFs evaluated at~$M/\sqrt{s}\,\exp (\pm Y)$, which corresponds to both high and low momentum fractions for a massive state at forward rapidities. Threshold resummation is most likely to constrain the PDFs at high momentum fractions, and so affect such an observable through both the PDFs and the partonic cross section. It is therefore prudent to study the effects of including threshold resummation in determinations of PDFs.

Several studies of resummation and its effects on the determinations of PDFs have been performed. One such study sought to estimate the effects that resummation in DIS would have on the valence quark PDFs~\cite{Sterman2000}. Another study used resummation in a PDF fit of DIS data to gauge the effects of threshold resummation on PDFs at large $x$~\cite{Corcella2005}. It was demonstrated in another study that pion PDFs determined using threshold resummation increased the accuracy of theoretical predictions over fixed order calculations~\cite{Aicher2010}. Recently, the NNPDF collaboration released a PDF set that included threshold resummation for both DIS and LPP in its determination~\cite{Bonvini2015}.
 
This study begins in \Secref{secResum} with a summary of the theory behind threshold resummation. The size and shape of threshold resummation corrections in DIS and LPP using current methods are reviewed in \Secref{secDISLPP}, followed by a description of the results of several PDF fits with resummation included in \Secref{secPDF}. An analysis of these results leads into a presentation of an enhanced approach to the calculation of the resummation exponent for the LPP rapidity and $x_{F}$ distributions in \Secref{secResum2}. After a study of the effect of using this enhanced resummation formalism in fitting PDFs, concluding remarks are made in \Secref{secConclude}.

\section{\label{secResum}Threshold Resummation}
In hadronic processes such as DIS and LPP, an arbitrarily large number of gluons may be emitted from the quarks and gluons taking part in the leading order (LO) reaction. These gluons can have negligibly small energies, resulting in a scattering that appears very similar kinematically to the LO process. A process with such a configuration of gluon energies is said to be in the threshold region.

It is well known that when the kinematics of a hadronic process is constrained it results in logarithms that become large at threshold. It is possible for these threshold logarithms to cause terms in the perturbative series to be as large as terms of a lower order; in such a situation, truncating the perturbative series at a fixed order can no longer be considered an accurate approximation of the full series. Threshold resummation is used to account for this by calculating the contributions of the threshold logarithms to all orders in perturbation theory.

The resummation calculation is usually performed in Mellin space via a Mellin transform of the observable. As an example, the Mellin transform of the DIS observable $F_{2}$ with respect to the Bjorken scaling variable $x$ and its inverse Mellin transform are given by
\begin{align}
\begin{split}
\tilde{F}_{2} (N) &= \int_{0}^{1} \mathrm{d}x \, x^{N-1} F_{2} (x)\quad\mathrm{and}\\
F_{2} (x) &= \int_{c-\I \infty}^{c+\I \infty} \mathrm{d}N \, x^{-N} \tilde{F}_{2} (N)\mathrm{,}\label{eqMELLIN}
\end{split}
\end{align}
respectively. Scale dependence has been suppressed in \Eqtextref{eqMELLIN} for simplicity, and will be suppressed throughout this study unless necessary.

The Mellin transform is a useful tool because the phase space of the soft gluons is a convolution integral in momentum space; a Mellin transform with respect to certain kinematic variables converts such a convolution integral into a simple product; the contribution of the soft gluons is therefore factorized in Mellin space (see, e.g., \Refcite{Catani1997}). A Mellin transform with respect to the Bjorken scaling variable $x$ factorizes soft gluon phase space of the structure functions in DIS, while the ratio $\tau=Q^{2}/s$ with lepton pair mass $Q$ and center of mass energy squared $s$ is used in the rapidity-integrated LPP differential cross section. There are methods that perform resummation directly in momentum space using soft collinear effective theory~\cite{Manohar2003,Idilbi2005,Becher2008}, but we will not address these in this study.

In Mellin space, the contribution of the threshold logarithms to all orders in perturbation theory is known to be an exponential. In DIS and the LPP rapidity-integrated mass distribution, the resummed partonic cross section takes the general form~\cite{Sterman1987a,Catani1989}
\begin{equation}
\tilde{C}^{Res}(N,\alphas)=C^{LO}(N,\alphas)g_{0}(\alphas)\exp\boldsymbol{(}G(N,\alphas)\boldsymbol{)}\mathrm{,}\label{eqRESUM}
\end{equation}
where $N$ is the Mellin moment, $C^{LO}(N,\alphas)$ is the LO term in Mellin space, $g_{0}(\alphas)$ is a perturbative series that collects threshold-enhanced terms independent of $N$, and $G(N,\alphas)$ contains the threshold logarithms. The calculation of the functions $g_{0}(\alphas)$ and $G(N,\alphas)$ can be found in, for example, Refs.~\cite{Sterman1987a,Catani1989,Catani1998}.

The accuracy of the resummed calculation depends on the power of the threshold logarithms included in $G(N,\alphas)$ relative to the strong coupling: for instance, at next-to-leading logarithmic accuracy (NLL), all logarithms of order $\alphas^{i}\ln^{2j}N$ with $i-1 \leq j \leq i$ must be included in $G(N,\alphas)$. Even if a high logarithmic accuracy is used in the resummation calculation, the resummation exponent itself only includes threshold-enhanced terms. Therefore, exclusively using resummation to calculate an observable loses information about kinematic regions away from threshold.

Adding a fixed order result to the resummed calculation retains the information that is not enhanced at threshold. However, the threshold terms are double-counted during this addition since they are present in both the fixed order calculation and resummation exponent. The double-counted or ``matching'' terms can be calculated via a series expansion of \Eqtextref{eqRESUM} around $\alphas=0$. For each order in the perturbative series that is included in the calculation, the equivalent order from the matching terms should be removed by subtracting them from the resummation exponent. We will be working at NLL accuracy matched to NLO in this study, abbreviated NLO$+$NLL.

In order to obtain meaningful results from the resummation formula, the calculation must be inverted back to momentum space through an inverse Mellin transform, as in \Eqtextref{eqMELLIN}. However, during the calculation of $G(N,\alphas)$ the running coupling is integrated below the Landau pole~\cite{Catani1996}.
As a consequence, the Landau pole appears in the resummation formula at $N_{L}=\exp\boldsymbol{(}1/(2b_{0}\alphas)\boldsymbol{)}$ for LPP and at $N_{L}^{2}$ for DIS, where
\begin{subequations}\label{eqALPHASBETA}
\begin{align}
\frac{\mathrm{d}\alphas}{\mathrm{d}\ln \mu^2} &= -b_{0}\alphas^2 (1+b_{1}\alphas + \ldots ) \label{eqALPHAS}\\
b_{0} &= \frac{11 C_{A} - 4 T_{R} n_{f}}{12\pi} = \frac{33-2n_{f}}{12\pi}\mathrm{,}\label{eqBETA0}\\
b_{1} &= \frac{17C_{A}^{2}-(10C_{A}+6C_{F})T_{R}n_{f}}{24\pi^{2}b_{0}} = \frac{153-19 n_{f}}{24\pi^{2}b_{0}}\mathrm{,}\label{eqBETA1}
\end{align}
\end{subequations}
and $n_{f}$ is the number of active quark flavors. Naively taking an inverse Mellin transform of the resummation formula would therefore include information from the Landau pole.

However, these are spurious effects and a prescription must be adopted in order to properly exclude them from the resummation calculation~\cite{Catani1996}. Many such prescriptions have been developed, but we use the minimal prescription (MP)~\cite{Catani1996} and the Borel prescription (BP)~\cite{Forte2006,Abbate2007} in this study. We use the form of the BP given by equation (3.29) in \Refcite{Bonvini2011}. The minimal and Borel prescriptions differ in ``subleading'' terms, which are terms that do not increase as rapidly as the threshold logarithms of the desired logarithmic accuracy. The difference between these prescriptions can therefore be interpreted as one estimate of the theoretical uncertainty associated with the resummation formalism~\cite{Bonvini2011}.

\section{\label{secDISLPP}Resummation in DIS and LPP}
\begin{figure}
		\centering
		\includegraphics{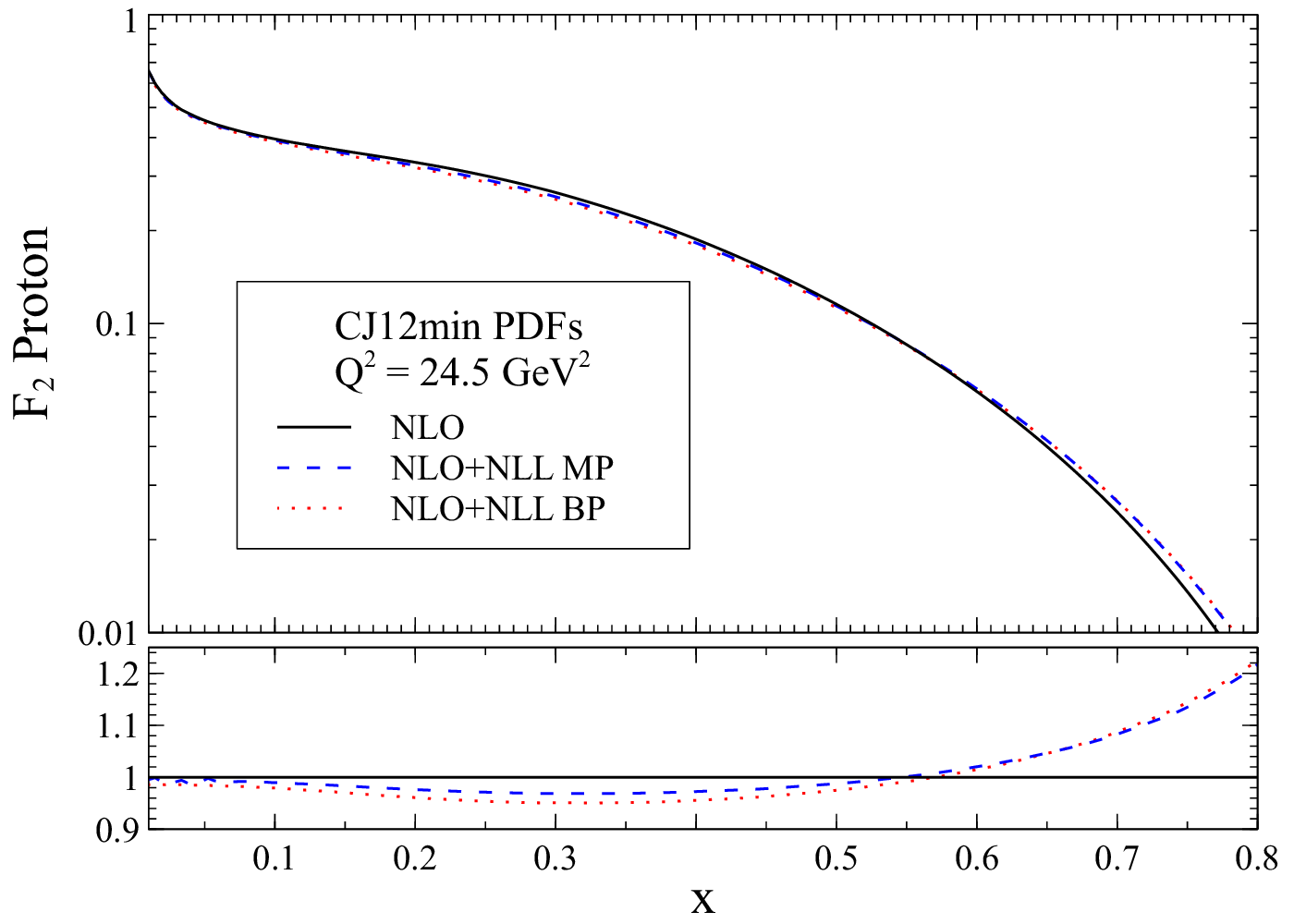}
		\caption{\label{figDIS245RESUM}A plot demonstrating resummation effects for $F_{2}$ at ${Q^{2}=24.5~\mathrm{GeV}^{2}}$ in electron-proton DIS. The upper plot shows the NLO (solid), NLO+NLL using MP (dashed), and NLO+NLL using BP (dotted) curves in a semi-log plot. The lower plot shows the same curves normalized to NLO.}
\end{figure}

There have been many studies on resummation in DIS~\cite{Sterman2000,Corcella2004,Corcella2005,Abbate2007}. Since the effects of DIS resummation are well understood, we will not go into great detail when examining their phenomenological implications. It has been shown that the effects of resummation in DIS are modest except at high $x$ and moderate $Q^{2}$~\cite{Sterman2000}, as demonstrated by \Figref{figDIS245RESUM} where the resummation corrections to the structure function $F_{2}$ in electron-proton DIS are a $20\%$ increase over NLO at $x\approx 0.8$. Since $F_{2}\sim x(4u_{v}+d_{v})$ at high $x$, with $u_{v}$ being the valence up quark PDF and $d_{v}$ being the valence down quark PDF, this effect will serve to decrease the up and down quark PDFs at high $x$. The effect of including DIS resummation on PDFs has already been estimated to be mild outside this kinematic region~\cite{Sterman2000}.

Resummation in the LPP rapidity distribution is a more recent development because the hadronic rapidity complicates the factorization of soft gluon phase space~\cite{Bolzoni2006,Mukherjee2006}. Unlike in DIS, where a single Mellin transform enables one to organize the threshold logarithms into an exponential, two integral transforms are required to do the same for the rapidity distribution. In methods currently employed in phenomenological studies, a Fourier transform with respect to rapidity is used in addition to the Mellin transform with respect to $\tau$. Performing a ``Mellin-Fourier'' transform and then organizing the threshold logarithms into an exponential can be difficult, but it was shown that by ignoring some subleading terms one can use the familiar rapidity-integrated resummation exponent~\cite{Mukherjee2006}:
\begin{equation}
\tilde{C}^{Res}(N,M,\alphas)\approx\int_{0}^{1}\mathrm{d}\hat{\tau}\,\hat{\tau}^{N-1}\cos\left(\frac{M}{2}\ln\frac{1}{\hat{\tau}}\right)C^{Res}(\hat{\tau},\alphas)\mathrm{,}\label{eqCOSINERES}
\end{equation}
where $M$ is the Fourier moment and $C^{Res}(\hat{\tau},\alphas)$ is the resummation formula for rapidity-integrated LPP in momentum space. The variable of integration $\hat{\tau}=\tau/(x_{1}x_{2})$ is the parton level analogue of the ratio $\tau$, with $x_{1}$ and $x_{2}$ being the momentum fractions of partons from the two hadrons involved in the reaction.

The cosine term in \Eqtextref{eqCOSINERES} is a source of ambiguity in the resummation calculation~\cite{Mukherjee2006}. One may choose to retain it by using
\begin{align}
\cos\left( \frac{M}{2}\ln\frac{1}{\hat{\tau}}\right) &= \frac{1}{2}(\hat{\tau}^{\frac{\I M}{2}}+\hat{\tau}^{-\frac{\I M}{2}})\mathrm{,}\label{eqCOSINEMETHOD}
\end{align}
thereby resulting in two terms in the resummation formula with shifted Mellin moments. Alternatively, one can recognize that 
\begin{align}
\cos\left( \frac{M}{2}\ln\frac{1}{\hat{\tau}}\right) &\approx 1+\mathcal{O}(M^{2}[1-\hat{\tau}]^{2})\approx 1\mathrm{,}\label{eqEXPANSIONMETHOD}
\end{align}
since the $\mathcal{O}(M^{2}[1-\hat{\tau}]^{2})$ terms are subleading~\cite{Mukherjee2006}. The method that uses \Eqtextref{eqCOSINEMETHOD} will be referred to as the ``cosine'' method while the method that uses \Eqtextref{eqEXPANSIONMETHOD} will be called the ``expansion'' method.

\begin{figure}[!ht]
		\centering
		\includegraphics{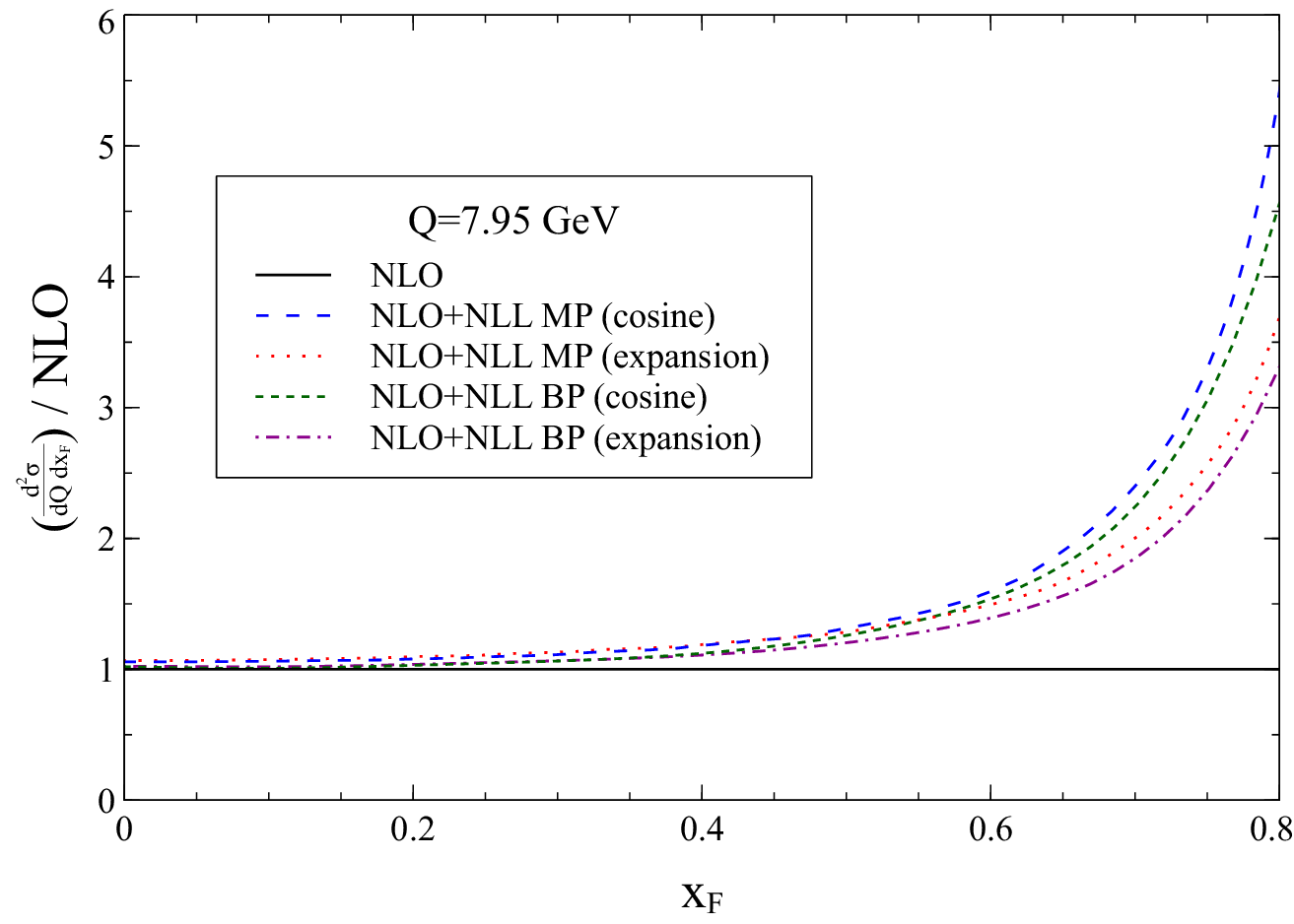}
\caption{\label{figLPPXF695NLONORM}A plot demonstrating the resummation effects on the proton-proton LPP $x_{F}$ distribution at $Q=7.95~\mathrm{GeV}$ and $\sqrt{s}=38.75$ GeV. The curves depict the calculations at NLO (solid), NLO+NLL using MP (dashed) and NLO+NLL using BP (fine dashed) with the cosine method, and NLO+NLL using MP (dotted) and NLO+NLL using BP (dot-dashed) with the expansion method, all normalized to NLO.}
\end{figure}

However, some LPP data sets do not measure data that is differential with respect to rapidity, but instead differential with respect to $x_F =2p_{L}/\sqrt{s}$, where $p_{L}$ is the longitudinal momentum of the lepton pair. Using LO kinematics, the rapidity can be related to $x_{F}$ using
\begin{equation}
Y=\frac{1}{2}\ln\frac{\sqrt{x_{F}^2+4\tau}+x_{F}}{\sqrt{x_{F}^2+4\tau}-x_{F}}\mathrm{.}\label{eqRAPXF}
\end{equation}
Since threshold occurs in the same kinematic region as the LO term, \Eqtextref{eqCOSINERES} is also valid for the $x_F$ distribution using \Eqtextref{eqRAPXF}, though there is an additional overall factor of $(x_{F}^2+4\tau)^{-\frac{1}{2}}$ associated with its LO term. The effects of the various resummation formulae for the $x_F$ distribution can be seen in \Figref{figLPPXF695NLONORM}. The resummation corrections are relatively moderate at low $x_{F}$, but at high $x_{F}$ they are very large: The resummation calculation is five times larger than NLO using the cosine method and three times larger using the expansion method at $x_{F}\approx 0.8$.

\section{\label{secPDF}PDF Fits Using Resummation: Cosine and Expansion Methods}
Using the methods outlined in \Secref{secResum} and \Secref{secDISLPP}, it is possible to perform global fits of PDFs that include threshold resummation in both DIS and LPP. Here we report on four different fits that were performed in order to account for the ambiguity present in the choice of resummation prescription (MP or BP) and the method used to calculate resummation corrections for the LPP $x_{F}$ distribution (cosine or expansion). These fits constrain the PDFs utilizing a collection of data that is similar to that of the CJ12 PDF sets~\cite{Accardi2010}: the DIS~\cite{Benvenuti1989,Benvenuti1990,Whitlow1992,Malace2009,Arneodo1997,Arneodo1997a,Aaron2010}, jet production~\cite{Affolder2001,Aaltonen2008,Abbott2001,Abazov2008}, and photon plus jet production data sets~\cite{Abazov2008a} are the same, but only the E866 proton-proton and proton-deuterium sets~\cite{Hawker1998,Webb2007,Reimer2006} are used for LPP. The same cuts that are used in the CJ12 PDF sets are used here, with $Q^{2} \geq 1.69~\mathrm{GeV}^{2}$ and $W^{2} \geq 3~\mathrm{GeV}^{2}$. In order to account for the nonperturbative effects that appear in DIS at lower $Q^{2}$ and $W^{2}$, the higher twist corrections, target mass corrections, and nuclear effects used in the CJ12min set are adopted. The PDF parameterization and choice of fit parameters that are allowed to vary, including the normalization parameters for each data set, are also patterned after the CJ12min set.

In addition to the standard CJ12 cuts, additional cuts are applied to some of the data sets. A cut of $x\geq 10^{-3}$ is applied to all DIS data, as resummation calculations below that region can become numerically unstable. Additionally, it was noted in \Refcite{Alekhin2006} that the E866 data at low values of $Q^{2}$ and high $x_{F}$ were not well described with an NNLO calculation using PDFs that were fit to DIS data. It was assumed in that study that this discrepancy arises from underestimated systematic uncertainties. Since threshold resummation includes the large threshold contributions from all orders, including NNLO, this trend can appear in these fits as well. Therefore, a cut $Q^{2}>32.5$ Ge$\mathrm{V}^{2}$ is applied to the E866 data sets in order to exclude the affected data.

\begin{figure}
		\centering
		\includegraphics{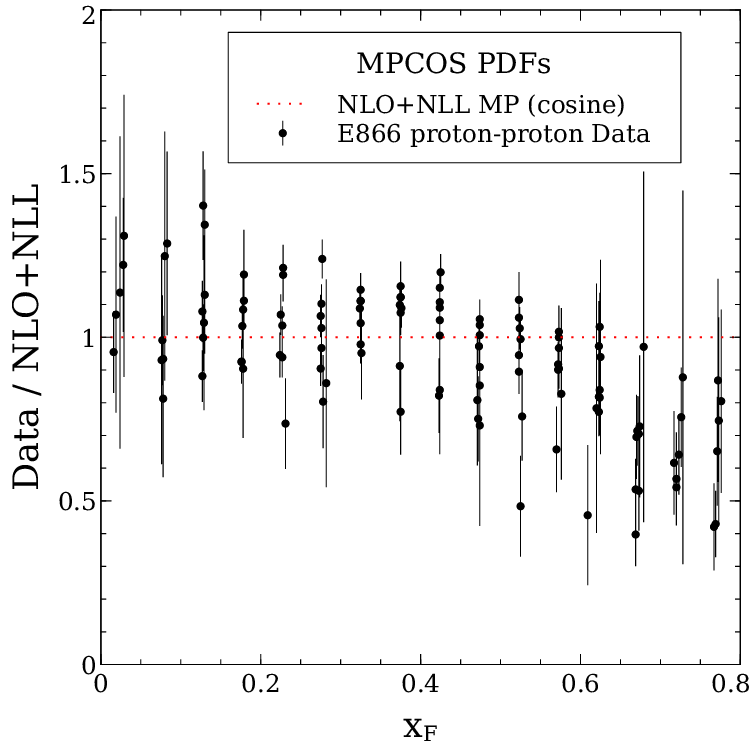}
		\caption{\label{figMPCOSFITCOMP}A plot of data points from the E866 proton-proton data at $\sqrt{s}=38.75$ GeV and $Q > 5.7$ GeV, with statistical and systematic errors added in quadrature. Note that some data points with exceptionally large uncertainties or that vary greatly from the other data have been excluded from the plot for the sake of aiding in visualization. The data have been normalized to a NLO+NLL calculation using the MP and cosine methods with the PDFs determined in the corresponding fit. Note that the NLO+NLL calculation was shifted by the normalization parameter from the fit, which resulted in an approximately $30\%$ reduction. The high $\chi^2$ is demonstrated by the fact that the resummed curve lies well above the data points at highest $x_F$, some of which have relatively small uncertainties.}
\end{figure}

\begin{table}[hbt]
\caption{\label{tabCHI2FOURFITS}The resultant $\chi^{2}$ values of the E866 data sets, as well as the combined $\chi^{2}$ of all the data sets, for the four fits using the cosine and expansion methods. The Total + Norm entries include the $\chi^2$ penalty incurred by the variation of the fitted normalization parameters from the published normalizations of the data sets.}
\begin{center}
\begin{tabular}{ l l c c c c} 
 \toprule
 {}               & {}            & \multicolumn{4}{c}{$\chi^{2}$}     \\
 \cline{3-6}
 Experiment       & No. points    & \;\; MP cosine  & \;\; BP cosine  & \;\; MP expansion  & \;\; BP expansion\\
 \cline{1-6}
 E866 (pp)        & 136           & 441.3      & 385.4      & 291.5         & 254.7  \\ 
 E866 (pd)        & 144           & 1056.6     & 924.7      & 573.5         & 471.4  \\ 
 \cline{1-6}
 Total            & 3655          & 4882.4     & 4720.8     & 4155.4        & 3984.7 \\
 Total + Norm     & {}            & 4970.6     & 4810.0     & 4207.1        & 4048.3 \\
 \botrule
\end{tabular}
\end{center}
\end{table}

The resultant $\chi^2$ values for the LPP data sets and overall $\chi^2$ values from each of the fits are summarized in \Tabref{tabCHI2FOURFITS}. The results of these fits exhibit a noticable trend: The LPP data is not described by the resultant PDFs for any of the fits, with $\chi^{2}$ per degree of freedom being well above expected levels. Additionally, the increase in $\chi^2$ of the Total~+~Norm row of \Tabref{tabCHI2FOURFITS} demonstrates that the fitted normalization parameters of the data sets do not line up with the published normalizations. To demonstrate why the $\chi^{2}$ is so large, \Figref{figMPCOSFITCOMP} shows E866 proton-proton data normalized to the NLO+NLL prediction using the PDFs fit with the MP and cosine method. At low $x_{F}$ the resummation calculation is largely consistent with data. However, much of the data at high $x_F$, which have relatively small errors, are well below the NLO+NLL calculation, despite the PDFs having been fit to this data. The region of high $x_{F}$ is where LPP resummation corrections are largest, suggesting that the cosine and expansion methods lose accuracy at high $x_{F}$ and increase too rapidly. In the recent NNPDF set that was fit using resummation, similar issues were noted: the high rapidity region was found to be problematic and therefore excluded from the fit~\cite{Bonvini2015}.

\section{\label{secResum2}Enhanced Resummation}
Since both the cosine and expansion methods, independent of prescription, increase too rapidly at high $x_F$, it is necessary to reformulate the resummation techniques used in the LPP rapidity and $x_F$ distributions. We now seek to resolve these problems by finding a resummation formalism that has smaller corrections at high $x_F$ and is more consistent with data. In order to accomplish this, we consider finding the resummation exponent for the $x_{F}$ distribution without ignoring certain terms that result in the formulation of \Eqtextref{eqCOSINERES}.

\subsection{\label{secEnhanced}Enhanced Resummation Formalism}

We suggest the following resummation formula valid up to NLL accuracy, first derived in \Refcite{Catani1989} in the DIS scheme but presented here in the $\overline{\mathrm{MS}}$ scheme:
\begin{align}
\begin{split}
G(N_{1},N_{2},\alphas)&=\int_{0}^{1}\!\mathrm{d}x_{1}\int_{0}^{1}\!\mathrm{d}x_{2}\frac{(x_{1}^{N_{1}-1}\!-\!1)(x_{2}^{N_{2}-1}\!-\!1)}{(1\!-\!x_{1})(1\!-\!x_{2})}A_{q}\boldsymbol{(}\alphas ([1\!-\!x_{1}][1\!-\!x_{2}]Q^{2})\boldsymbol{)}\\
&\quad{}+\int_{0}^{1}\mathrm{d}x\,\frac{(x^{N_{1}-1}-1)+(x^{N_{2}-1}-1)}{1-x}\int_{\muf^{2}}^{(1-x)Q^{2}}\frac{\mathrm{d}k^{2}}{k^{2}}A_{q}\boldsymbol{(}\alphas(k^{2})\boldsymbol{)}\mathrm{,}\label{eqIMPROVEDRESUM}
\end{split}
\end{align}
where $N_{1,2}$ are the Mellin conjugates of $x_{1,2}^{0}=\sqrt{\tau}\mathrm{e}^{\pm Y}$ in the rapidity distribution and $x_{1,2}^{0}=\frac{1}{2}(\sqrt{x_{F}^{2}+4\tau}\pm x_{F})$ in the $x_{F}$ distribution. The function $A_{q}(\alphas)$ is a perturbative expansion in $\alphas$:
\begin{subequations}\label{eqA1A2}
\begin{align}
A_{q}(\alphas)&=\sum_{i=1}^{\infty}\alphas^{i}\, A_{i}\mathrm{,}\label{eqRESFUNCPERTEXP}\\
A_{1}&=\frac{C_{F}}{\pi}\mathrm{,}\quad\mathrm{and}\label{eqRESUMCONSTA1}\\
A_{2}&=\frac{C_{F}}{2\pi^{2}}\left[C_{A}\left(\frac{67}{18}-\frac{\pi^{2}}{6}\right)-\frac{10}{9}T_{R}n_{f}\right]\mathrm{.}\label{eqRESUMCONSTA2}
\end{align}
\end{subequations}
The resummation exponent of \Eqtextref{eqIMPROVEDRESUM} is incomplete: another function, $D(\alphas)$, must also be present in the resummation formalism. $D(\alphas)$ is a process-dependent perturbative expansion similar in nature to $A_{q}$ that contains information from wide-angle gluon emissions~\cite{Moch2005}; however, since $D_{1}=0$ in LPP~\cite{Catani1989} its exclusion from \Eqtextref{eqIMPROVEDRESUM} is still accurate as long as we remain at NLL accuracy. $G(N_{1},N_{2},\alphas)$ is to be used in a function similar to that of \Eqtextref{eqRESUM}, given by
\begin{align}
\tilde{C}^{Res}(N_{1},N_{2},\alphas) &= \tilde{C}^{LO}(N_{1},N_{2},\alphas)g_{0}(\alphas)\exp\boldsymbol{(}G(N_{1},N_{2},\alphas)\boldsymbol{)}\mathrm{.}\label{eqENHANCERESUM}
\end{align}
The resummation method using \Eqtextref{eqENHANCERESUM} will be referred to as the ``enhanced'' resummation formalism in this study. The function $g_{0}(\alphas)$ to be used with \Eqtextref{eqENHANCERESUM} is the same as that used in the resummation formula of the LPP mass distribution and is found in, e.g., Refs.~\cite{Moch2005,Bonvini2010}.

Following the methodology of \Refcite{Vogt2001a}, we have computed the integrals in the exponential of \Eqtextref{eqIMPROVEDRESUM} to NLL and have found~\cite{Westmark2015}:
\begin{subequations}\label{eqIMPROVEEXP}
\begin{align}
G(N_{1},N_{2},\alphas)&=(\ln N_{1}+\ln N_{2})g_{1}(\lambda)+g_{2}(\lambda)+\mathcal{O}(\alphas \,\lambda)\mathrm{,}\label{eqGEXPDYFULL}\\
g_{1}(\lambda)&=\frac{A_{1}}{b_{0}\lambda}[(1-\lambda)\ln(1-\lambda) +\lambda ]\mathrm{,}\label{eqG1DYFULL}\\
\begin{split}
g_{2}(\lambda)&=\frac{A_{1}b_{1}}{b_{0}^{2}}[\ln (1-\lambda )+\frac{1}{2}\ln^{2}( 1-\lambda)-\lambda]\\
&\quad{}-\frac{2A_{1}\gamma_{E}}{b_{0}}\ln(1-\lambda)+\frac{A_{1}}{b_{0}}\ln(1-\lambda)\ln\frac{Q^{2}}{\mur^{2}}\\
&\quad{}-\frac{A_{2}}{b_{0}^{2}}[\lambda+\ln(1-\lambda)]+\frac{A_{1}}{b_{0}}\lambda\ln\frac{\muf^{2}}{\mur^{2}}\mathrm{,}
\end{split}\label{eqG2DYFULL}
\end{align}
\end{subequations}
where
\begin{align}
\lambda &= \alpha_{S}b_{0}(\ln N_{1}+\ln N_{2})\mathrm{,}\label{eqLAMBDA}
\end{align}
$b_{0}$ and $b_{1}$ are the coefficients of the beta function of $\alpha_{S}$ given in \Eqtextref{eqALPHASBETA}, $\muf$ and $\mur$ are the factorization and renormalization scales, $\alphas\equiv \alphas(\mur^{2})$, and the constants $A_1$ and $A_2$ are given by \Eqtextref{eqA1A2}. The $\mathcal{O}(\alphas)$ expansion of \Eqtextref{eqENHANCERESUM}, used in the matching procedure to remove double-counting terms, is given by
\begin{align}
\begin{split}
\alphas\left. \frac{\partial C^{Res}(N_{1},N_{2},\alphas)}{\partial\alphas}\right|_{\alphas =0}&=\frac{\alphas}{2\pi}C_{F}\Big\{ \ln\frac{Q^{2}}{\muf^{2}}[3-4\GAMMA-2(\ln N_{1}+\ln N_{2})]+\frac{4\pi^{2}}{3}\\
&\quad{}-8+4\GAMMA^{2}+[\ln N_{1}+\ln N_{2}]^{2}+4\GAMMA [\ln N_{1}+\ln N_{2}]\Big\}\mathrm{.}\label{eqMATCHING}
\end{split}
\end{align}
This result also acts as a check on the enhanced resummation formula, as it reproduces the large $N_{1}$, $N_{2}$ limit of the NLO calculation in double-Mellin space~\cite{Westmark2015}.

Additionally, the enhanced resummation formalism, \Eqtextref{eqENHANCERESUM}, is a general form of both the cosine and expansion methods. As should be apparent from applying \Eqtextref{eqCOSINEMETHOD} to \Eqtextref{eqCOSINERES}, the Mellin transform of the variable $\hat{\tau}$ is shifted by the Fourier moment $M$. The two shifted moments $N\pm \I M/2$ are exactly the two Mellin variables $N_{1}$ and $N_{2}$ present in \Eqtextref{eqIMPROVEDRESUM}. Therefore, through a simple comparison of \Eqtextref{eqIMPROVEEXP} to the widely-known rapidity-integrated exponent it can be seen that the cosine method is equivalent to evaluating
\begin{align}
\begin{split}
\tilde{C}^{Res}_{cosine}(N,M,\alphas) &= \frac{1}{2}[\tilde{C}^{Res}(N-\frac{\I M}{2},N-\frac{\I M}{2},\alphas)+\tilde{C}^{Res}(N+\frac{\I M}{2},N+\frac{\I M}{2},\alphas)]\\
 &= \frac{1}{2}[\tilde{C}^{Res}(N_{1},N_{1},\alphas)+\tilde{C}^{Res}(N_{2},N_{2},\alphas)]\mathrm{,}\label{eqCOSINEGENERAL}
\end{split}
\end{align}
where $\tilde{C}^{Res}_{cosine}$ is the cosine method's resummation formula and $\tilde{C}^{Res}$ is the resummation formula of \Eqtextref{eqENHANCERESUM}. Similarly, the expansion method is equivalent to using
\begin{align}
\tilde{C}^{Res}_{expansion}(N,M,\alphas) = \tilde{C}^{Res}(N,N,\alphas) = \tilde{C}^{Res}(\frac{1}{2}[N_{1}+N_{2}],\frac{1}{2}[N_{1}+N_{2}],\alphas )\mathrm{.}\label{eqEXPANSIONGENERAL}
\end{align}

\subsection{\label{secPrescriptions}Resummation Prescriptions}

Since the Landau pole is still present in the enhanced resummation exponent, a resummation prescription must be applied. However, there are some nuances to using the resummation prescriptions with the enhanced resummation formalism. One of the consequences of the resummation exponent existing in double-Mellin space is that the Landau pole no longer corresponds to a single Mellin moment. Instead, the Landau pole in double-Mellin space occurs on the curve
\begin{equation}
N_{1}N_{2}=\exp \Big(\frac{1}{\alphas b_{0}}\Big)\mathrm{.}\label{eqLANDAUDOUBLEMELLIN}
\end{equation}

In the MP, the path of the inverse Mellin transform of the resummation exponent must be to the left of the Landau pole but to the right of all other poles~\cite{Catani1996}. However, the exact resummation exponent depends on two Mellin variables, and therefore two inverse Mellin transforms must be taken. The paths of these inverse Mellin transforms must be to the right of the poles in both $N_{1}$ and $N_{2}$ space, but performed in such a way that the Landau pole lies to the right of the paths.

It is apparent from \Eqtextref{eqLANDAUDOUBLEMELLIN} that $N_{1}$ is not independent from $N_{2}$ if the path of the inverse Mellin transform is to lie to the left of the Landau pole. For every fixed value of $N_{1}$, the Landau pole appears in $N_{2}$ space as
\begin{equation}
\widetilde{N}_{L}(N_{1})=\frac{1}{N_{1}}\exp \Big(\frac{1}{\alphas b_{0}}\Big)\mathrm{.}\label{eqLANDAUDBLEMELN2}
\end{equation}
Since $N_{1}$ takes complex values during the inverse Mellin transform, the Landau pole moves from the real $N_{2}$ axis into complex $N_{2}$ space. Therefore, for a fixed $N_{1}$ the path of the inverse Mellin transform with respect to $N_{2}$ must be carefully chosen to cross the real axis to the right of all other poles in $N_{2}$ space, but deformed such that it still remains to the left of the Landau pole in complex $N_{2}$ space.

The derivation of the BP for the enhanced resummation formalism can be performed in analogue to the derivation of \Refcite{Bonvini2011}. This is done by noting that the enhanced resummation exponent and matching terms are functions of $(\ln N_{1}+\ln N_{2})$ only. Taking care to account for the two inverse Mellin transforms, we arrive at the final form~\cite{Westmark2015}
\begin{align}
\begin{split}
C^{BP}(x_{1},x_{2})&\! =\! \frac{1}{(2\pi\I)^{2}}\!\oint\!\frac{\mathrm{d}\xi_{1}}{\xi_{1}\Gamma(\xi_{1})}\,\oint\!\frac{\mathrm{d}\xi_{2}}{\xi_{2}\Gamma(\xi_{2})}\,\left[\left(1-x_{1}\right)^{\xi_{1}-1}\right]_{+}\! \left[\left(1-x_{2}\right)^{\xi_{2}-1}\right]_{+}\\
&\quad{}\times x_{1}^{-\frac{\xi_{1}}{2}}\! x_{2}^{-\frac{\xi_{2}}{2}}\int_{0}^{c}\mathrm{d}t\,\E^{-t}\frac{\xi_{2}\tilde{C}^{BP}\left(\frac{\alphas b_{0}t}{\xi_{1}}\right)-\xi_{1}\tilde{C}^{BP}\left(\frac{\alphas b_{0}t}{\xi_{2}}\right)}{\xi_{2}-\xi_{1}}\mathrm{,}\label{eqDBLEMELLBPFINAL}
\end{split}
\end{align}
where the ``+'' notation represents the standard plus distributions:
\begin{align}
\int_{0}^{1}\mathrm{d}x\, \big[ (1-x)^{\xi-1}\big]_{+} f(x) = \int_{0}^{1}\mathrm{d}x\, (1-x)^{\xi-1} (f(x)-f(1))\mathrm{.}\label{eqPLUSDIST}
\end{align}
The function $\tilde{C}^{BP}$ in \Eqtextref{eqDBLEMELLBPFINAL} is the enhanced resummation exponent and matching terms as a function of $\lambda$ like in \Eqtextref{eqIMPROVEEXP}, $C^{BP}$ is the enhanced resummation exponent and matching terms in momentum space, and $c$ is a cutoff for the Borel integral~\cite{Bonvini2011}.

\subsection{\label{secPhenom}Phenomenology}

\begin{figure}
		\centering
		\includegraphics{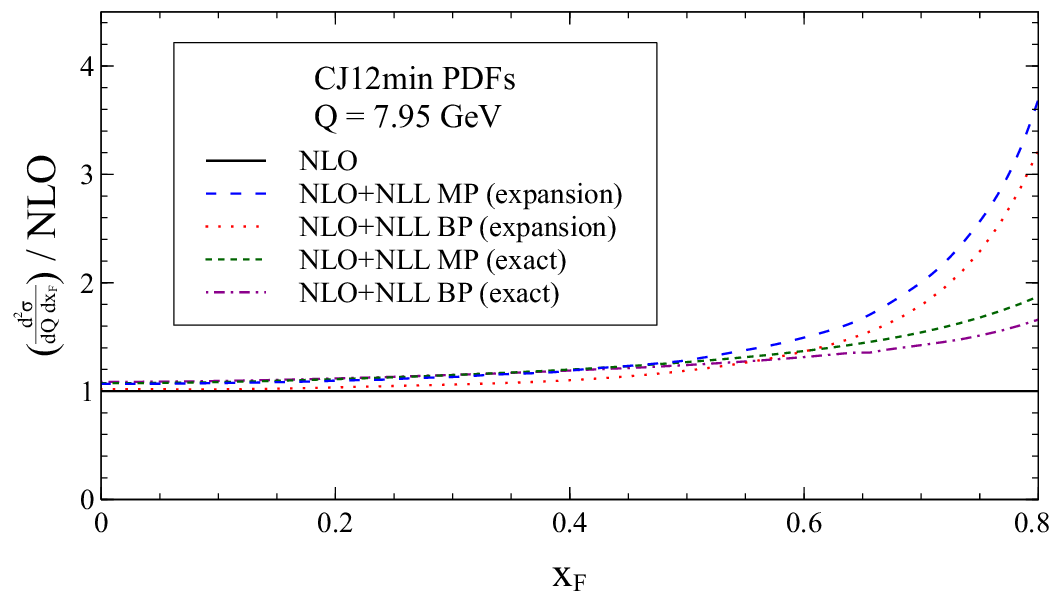}
		\caption{\label{figEXACTCOMPAREEXPAND}A plot showing the resummation corrections to the LPP $x_F$ distribution using the enhanced resummation formalism. The curves for NLO (solid), NLO+NLL using the MP and expansion method (dashed), NLO+NLL using the BP and expansion method (dotted), and NLO+NLL using the enhanced resummation formalism for both MP (fine dashed) and BP (dot-dashed) are all calculated at ${Q=7.95}$~GeV using CJ12min PDFs and are normalized to NLO.}
\end{figure}

In \Figref{figEXACTCOMPAREEXPAND}, we show the resummation corrections for the LPP $x_F$ distribution normalized to NLO at $Q=7.95$ GeV and $\sqrt{s}=38.75$ GeV using the expansion method and the improved resummation formalism~\cite{Westmark2015}. The corrections from improved resummation are of approximately the same size as those of the expansion method at low $x_F$, but at high $x_F$ the improved resummation formalism increases significantly less rapidly, being less than twice the size of NLO at high $x_F$. In addition, the difference between the MP and BP is slightly less pronounced when using the improved resummation formalism.

\Figref{figEXACTCOMPAREEXPAND} also affirms the approach used in the recent NNPDF set determined using threshold resummation. In that study, the expansion method was employed and the high $x_F$ region was excluded because it was a source of discrepancy~\cite{Bonvini2015}. \Figref{figEXACTCOMPAREEXPAND} demonstrates that the expansion method is similar to the more general enhanced resummation method in the low $x_F$ region included in the fit, but deviates strongly in the high $x_F$ region.

\subsection{\label{secPDF2}Determinations of PDFs}

We next present the results of two global PDF fits using the enhanced resummation formalism~\cite{Westmark2015}, similar to those of \Secref{secPDF}. A brief summary of the results of these fits can be found in \Tabref{tabCHI2TWOFITS}. The $\chi^{2}$ results are presented for the LPP data sets used in the fits, as well as the global $\chi^{2}$.

The quality of these two PDF fits is comparable to that of PDFs determined using NLO. It should be noted that the normalization parameters of the LPP data sets are well above one, being around $1.2$ for the proton-proton and proton-deuterium E866 data in both fits. We have noted in an ad hoc global fit that using NNLO+NNLL reduces these normalization parameters so that they are closer to 1, implying that the higher normalization parameters are a coincidence of the resummation formalism at NLO+NLL. However, no firm conclusions can be drawn until additional data with low uncertainties in the high $x_{F}$ region become available from other experiments, such as E906/SeaQuest~\cite{Reimer2006a}.

\begin{table}[hbt]
\begin{center}
\caption{\label{tabCHI2TWOFITS}The resultant $\chi^{2}$ values of the E866 data sets, as well as the combined $\chi^{2}$ of all the data sets, for the two fits using the improved resummation formalism. The Total + Norm entries include the $\chi^2$ penalty incurred by the variation of the fitted normalization parameters from the published normalizations of the data sets.}
\begin{tabular}{ l l c c } 
 \toprule
 {}               & {}         & \multicolumn{2}{c}{$\chi^{2}$}     \\
 \cline{3-4}
 {}               & No. points & MP     & BP \\
 \cline{1-4}
 E866 (pp)        & 136        & 176.0  & 163.1    \\ 
 E866 (pd)        & 144        & 219.0  & 169.8    \\ 
 \cline{1-4}
 Total        & 3655       & 3635.1 & 3530.6  \\
 Total + Norm & {}         & 3674.2 & 3580.6  \\
 \botrule
\end{tabular}
\end{center}
\end{table}

\begin{figure}
		\centering
		\includegraphics{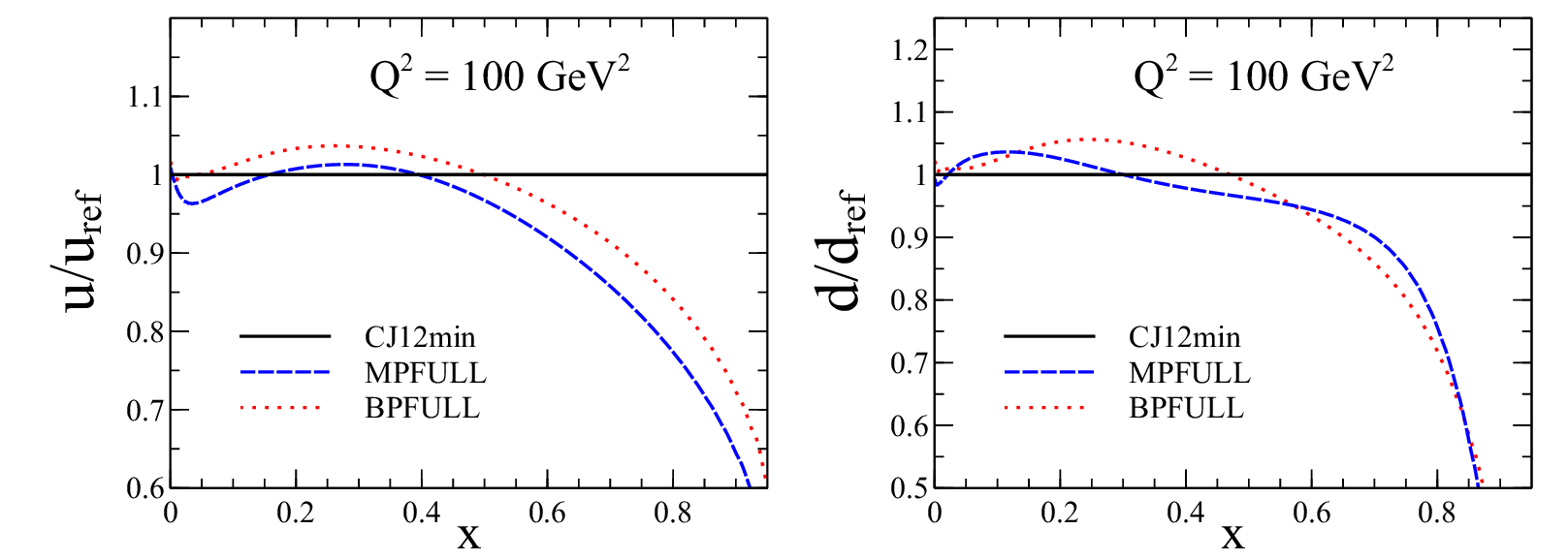}
		\caption{\label{figMPBPFULLPDFSHAPES}A comparison of the shapes of the up and down quark distributions from the CJ12min PDFs to the two fitted PDF sets at ${Q^{2}=100~\mathrm{GeV}^{2}}$. Both distributions are both shown normalized to the CJ12min PDFs, which act as the reference set of PDFs.}
\end{figure}

As a final point, \Figref{figMPBPFULLPDFSHAPES} shows the shapes of the PDFs that result from these two fits compared to those from the CJ12min NLO PDF set. The fitted PDFs exhibit an approximately $20\%$ reduction in the up quark distribution and $30\%$ reduction in the down quark distribution at $x\approx 0.8$. This is to be expected for the up and down quark distributions, since DIS resummation corrections are a $20\%$ increase over NLO in this kinematic region and the DIS data included in the fit provides the strongest constraints to these distributions. It should be noted that in the regions where data provided strong constraints, the sea quark and gluon PDFs saw a less significant change of $\lesssim 10\%$~\cite{Westmark2015}.

\section{\label{secConclude}Conclusions}
We have presented a discussion of the effect of threshold resummation in the determination of parton distribution functions. In the course of this study, the currently available methods for threshold resummation in the LPP rapidity and $x_F$ distributions were demonstrated to disagree with LPP data at high $x_F$ due to the rapidly increasing size of the resummation corrections. In response to this, an existing resummation formalism was employed that does not make the same approximations as the current formalism and is consistent with LPP data over all $x_{F}$. Methods for using this formalism in the minimal and Borel prescriptions were derived and used in two global fits to demonstrate resummation effects on PDFs.

The results of this study lead to several conclusions. It can be inferred that the current methods of performing threshold resummation for the LPP rapidity and $x_{F}$ distributions are accurate approximations at low $x_{F}$ or rapidity, but lose validity at higher values of $x_{F}$ and rapidity due to the presence of a rapid increase in size that is incompatible with data. The enhanced resummation formalism was shown to be consistent with data in the high $x_{F}$ region, and in order to avoid overestimating the size of the resummation corrections should therefore should be used instead of the cosine or expansion methods at high rapidities or $x_{F}$. Additionally, comparisons were drawn to the NNPDF resummed fit, and the methods used in that study were validated; however, those methods exclude important information from the high rapidity or $x_F$ region. Finally, a determination of the PDFs that includes threshold resummation has a visible effect on the size and shape of the resultant PDFs, with the up and down quark distributions seeing an approximately $20\%$ and $30\%$ reduction at high $x$, respectively.

\bibliography{bibliography}
\end{document}